\documentclass[11pt]{article}
\usepackage{natbib}
\usepackage[english]{babel}
\usepackage[latin1]{inputenc}
\usepackage{lmodern}
\usepackage[T1]{fontenc}
\usepackage{amssymb,amsmath,amsthm,latexsym,amsfonts,amscd,dsfont,enumerate,bbm,mathabx}
\usepackage{graphicx,tikz}
\usepackage{listings}
\usepackage{booktabs}
\usepackage{mathtools}
\usepackage[gen]{eurosym}
\usepackage{algorithm}
\usepackage{algpseudocode}
\usepackage{soul}

\usepackage{pgfplots}
\pgfplotsset{compat=newest}

\definecolor{mylightyellow}{rgb}{1,1,.8}
\definecolor{mylightgreen}{rgb}{.8,1,.8}
\definecolor{mydarkred}{RGB}{178,34,34}
\definecolor{mydarkgreen}{RGB}{34,139,34}
\definecolor{mydarkblue}{RGB}{72,61,139}
\definecolor{mydarkyellow}{RGB}{218,165,32}

\definecolor{myblueA}{RGB}{52,41,39}
\definecolor{myblueB}{RGB}{92,81,109}
\definecolor{myblueC}{RGB}{132,121,179}
\definecolor{myblueD}{RGB}{172,161,249}
\definecolor{myblueE}{RGB}{212,201,255}

\usetikzlibrary{arrows,patterns,shapes,snakes,shadows,fit,backgrounds,positioning,decorations.pathmorphing,external}
\tikzstyle{nameusd} = [circle, draw, top color=white, bottom color=mydarkblue!50, draw=mydarkblue!75!black!100, drop shadow, minimum height=4em]
\tikzstyle{nameeur} = [circle, draw, top color=white, bottom color=mydarkred!50, draw=mydarkred!75!black!100, drop shadow, minimum height=4em]
\tikzstyle{collusd} = [rectangle,fill=mydarkblue!10, inner sep=0.2cm, rounded corners=5mm]
\tikzstyle{colleur} = [rectangle,fill=mydarkred!10, inner sep=0.2cm, rounded corners=5mm]
\tikzstyle{fixto} = [draw, -latex']
\tikzstyle{fixfrom} = [draw, latex'-]
\tikzstyle{floatto} = [draw, snake=coil, segment aspect=0, line before snake=2ex, line after snake=1ex, -latex']
\tikzstyle{floatfrom} = [draw, snake=coil, segment aspect=0, line before snake=2ex, line after snake=1ex, latex'-]
\tikzstyle{investor} = [rectangle, draw, top color=white, bottom color=mydarkyellow!50, draw=mydarkyellow!75!black!100, drop shadow, rounded corners, minimum height=3em, text width=4em, text centered]
\tikzstyle{trader} = [circle, draw, top color=white, bottom color=blue!30, draw=blue!50!black!100, drop shadow, minimum height=4em]
\tikzstyle{bank} = [rectangle, draw, top color=white, bottom color=red!20, draw=red!50!black!100, drop shadow, rounded corners, minimum height=3em, text width=4em, text centered]
\tikzstyle{market} = [rectangle, draw, top color=white, bottom color=green!20, draw=green!50!black!100, drop shadow, rounded corners, minimum height=3em, text width=4em, text centered]
\tikzstyle{yellowbox} = [rectangle, draw, top color=white, bottom color=mydarkyellow!50, draw=mydarkyellow!75!black!100, drop shadow, rounded corners, text centered]
\tikzstyle{fadedyellowbox} = [rectangle, draw=gray, text=gray, top color=white, bottom color=mydarkyellow!25, draw=mydarkyellow!50, drop shadow, rounded corners, text centered]
\tikzstyle{redbox} = [rectangle, draw, top color=white, bottom color=mydarkred!50, draw=mydarkred!75!black!100, drop shadow, rounded corners, text centered]
\tikzstyle{fadedredbox} = [rectangle, draw=gray, text=gray, top color=white, bottom color=mydarkred!25, draw=mydarkred!50, drop shadow, rounded corners, text centered]
\tikzstyle{bluebox} = [rectangle, draw, top color=white, bottom color=mydarkblue!50, draw=mydarkblue!75!black!100, drop shadow, rounded corners, text centered]
\tikzstyle{fadedbluebox} = [rectangle, draw=gray, text=gray, top color=white, bottom color=mydarkblue!25, draw=mydarkblue!50, drop shadow, rounded corners, text centered]
\tikzstyle{greenbox} = [rectangle, draw, top color=white, bottom color=mydarkgreen!50, draw=mydarkgreen!75!black!100, drop shadow, rounded corners, text centered]
\tikzstyle{fadedgreenbox} = [rectangle, draw=gray, text=gray, top color=white, bottom color=mydarkgreen!25, draw=mydarkgreen!50, drop shadow, rounded corners, text centered]

\pgfplotstableread{calibration_ngttf.dat}\calibngttf
\pgfplotstableread{cso_TTF.dat}\csottf
\pgfplotstableread{mco_TTF.dat}\mcottf
\pgfplotstableread{vola_deliveries.dat}\vdttf
\pgfplotstableread{DA_backbone.dat}\volada
\pgfplotstableread{swing_fixed_strike.dat}\swingkfix
\pgfplotstableread{swing_forward_start.dat}\swingkfwd

\usepackage[]{hyperref}
 \hypersetup{
 colorlinks=true,breaklinks=true,
 urlcolor= mydarkblue,linkcolor= mydarkblue,citecolor= mydarkgreen,
 pdfauthor={Nastasi, E.,Pallavicini, A.,Sartorelli, G.}
}

\newcommand{\Eq}[1]{{\begin{equation}{#1}\end{equation}}}

\newcommand{\Exo}[1]{\mathbb{E}\!\left[\,#1\,\right]}
\newcommand{\Ex}[2]{\mathbb{E}_{#1}\!\left[\,#2\,\right]}

\newcommand{\ExCo}[2]{\mathbb{E}\!\left[\left.\,#1\,\right|\,#2\,\right]}

\newcommand{\QxC}[2]{\mathbb{Q}\!\left\{\,#1\,\left.\right|\,#2\,\right\}}

\newcommand{\ind}[1]{1_{\{#1\}}}
\newcommand{\clip}[3]{\operatorname{clip}\!\left({#1},{#2},{#3}\right)}

\usepackage{eurosym}

\title{Pricing Commodity Swing Options\thanks{We thank Edoardo Vittori for introducing us to reinforcement learning algorithms.}}
\author{
Roberto Daluiso\thanks{Banca IMI Milan, {\tt roberto.daluiso@bancaimi.com}}
\and
Emanuele Nastasi\thanks{Exprivia, {\tt emanuele.nastasi@exprivia.com}}
\and
Andrea Pallavicini\thanks{Imperial College London and Banca IMI Milan, {\tt a.pallavicini@imperial.ac.uk}}
\and
Giulio Sartorelli\thanks{Banca IMI Milan, {\tt giulio.sartorelli@bancaimi.com}}
}
\date{
\small First Version: October 15, 2019.  This version: \today
}

\begin{document}

\maketitle

\begin{abstract}

In commodity and energy markets swing options allow the buyer to hedge against futures price fluctuations and to select its preferred delivery strategy within daily or periodic constraints, possibly fixed by observing quoted futures contracts. In this paper we focus on the natural gas market and we present a dynamical model for commodity futures prices able to calibrate liquid market quotes and to imply the volatility smile for futures contracts with different delivery periods. We implement the numerical problem by means of a least-square Monte Carlo simulation and we investigate alternative approaches based on reinforcement learning algorithms.

\end{abstract}

\bigskip

\noindent {\bf AMS classification codes:} 65C05, 91G20, 91G60.

\newpage
{\small \tableofcontents}
\vfill
{\footnotesize \noindent The opinions here expressed  are solely those of the authors and do not represent in any way those of their employers.}
\newpage

\pagestyle{myheadings} \markboth{}{{\footnotesize  Daluiso, Nastasi, Pallavicini, Sartorelli, Pricing Commodity Swing Options}}

\section{Introduction}
\label{sec:introduction}

In energy markets, a class of commonly traded contracts allows the buyer to select its preferred delivery strategy within daily or periodic constraints, while the purchase price can be fixed at inception or determined before the starting date of the delivery period by observing the prices of quoted futures contracts. These contracts are usually known as swing options since the buyer is allowed to swing  between a lower and an upper boundary in the commodity flow.

From the modeling point of view, the daily selection of the delivery strategy along with constraints on the total consumption force us to describe the swing option pricing problem as a specific type of a stochastic control problem for the optimal consumption strategy. The first works in the literature date back to the nineties and they focus on specific payoffs, see for instance \cite{Thompson1995}. The first contribution describing general swing option payoffs is \cite{Jaillet2004}, where the authors provide an efficient valuation framework and propose a stochastic process appropriate for energy prices. Alternative numerical approximations can be found in \cite{Haarbrucker2009}, \cite{Zhang2013}, \cite{Kirkby2020}. Investigations on the price dynamics of the underlying commodity can be found in \cite{Benth2012}, or in \cite{Eriksson2013} where L\'evy models are introduced.

The theoretical aspects of the stochastic control problem are described in \cite{Barrera2006}, where the delivery strategy is analyzed also by using neural networks, and later in \cite{Carmona2008} and \cite{Bardou2009}. In these papers a specific consumption strategy, named bang-bang, is discussed. According to this strategy only the minimum or maximum consumption allowed by all the constraints is selected on each delivery day. In particular in \cite{Bardou2009} sufficient conditions for the existence of an optimal bang-bang strategy are derived.

Our contribution to the literature is twofold. First, we propose a simple diffusive model for commodity futures prices, which is able to describe the volatility smile quoted by the market for futures contracts with different delivery periods. Our proposal starts from the extension of the local-volatility linear model presented in \cite{Nastasi2018}. We stress the importance of modelling futures prices with heterogeneous delivery periods since swing option prices depend both on day-ahead prices through the consumption strategy and on longer period futures contracts (usually one-month contracts) to determine the purchase strike prices. We also show how spikes can be included in our framework. Second, we investigate reinforcement learning algorithms as possible alternative to least-square Monte Carlo simulations. Here, we consider the proximal-policy optimization algorithm proposed in \cite{Schulman2017}, and we implement it in our specific case.

\medskip

The paper is organized as follows. In Section~\ref{sec:smile} we present the model we use to describe the prices of futures contracts on different delivery periods. Then, in Sections~\ref{sec:numerics} we present numerical examples derived by means of least-square Monte Carlo techniques. In particular we check the possibility of optimal bang-bang strategies to test the soundness of the numerical algorithm. We conclude the paper with Section~\ref{sec:rl} devoted to the application of reinforcement learning to swing option pricing.

\section{Modelling Commodity Smiles}
\label{sec:smile}

The local-volatility linear model presented in \cite{Nastasi2018} allows to describe futures prices in a parsimonious way while preserving a perfect fit to plain vanilla options quoted in the commodity market. Moreover, mid-curve options and calendar spread options can be calibrated by means of a best-fit procedure. In the original paper some extensions are discussed to introduce multiple risk factors to drive the curve dynamics and to allow for stochastic volatilities. Here, we stick to the one-dimensional specification of the model and we investigate how to extend it to deal with futures contracts on different delivery periods and to incorporate spikes.

\subsection{Calibration of Futures Option Smile}
\label{sec:calibration}

We start by considering futures contracts with the same delivery period (e.g.\ one month). We model their prices by introducing the price process $S_t$ of a rolling futures contract which can be identified with futures contracts quoted in the market on their last trading date (or on their first notification date if it occurs before the last trading date). We can think of this process as a ``fictitious'' spot price. We model the spot price by means of the process
\Eq{
s_t := \frac{S_t}{F_0(t)}
}%
where $F_0(t)$ is the futures price term structure as seen today. We select a local-volatility model with a linear drift for the process $s_t$, namely we write
\Eq{
ds_t = a(t) (1 - s_t) \,dt + \eta(t,s_t) s_t \,dW_t
\,,\quad
s_0 = 1
}%
where $W_t$ is a standard Brownian motion under the risk-neutral measure, $a(t)$ is a positive function of time, $\eta(t,s_t)$ is Lipschitz in the second argument, bounded and positive. With these assumptions the previous SDE has a unique positive solution for any time $t>0$.

We briefly summarize the results of \cite{Nastasi2018} which we employ in the following of the paper. First, we derive the dynamics followed by futures prices $F_t(T)$. We obtain
\Eq{
dF_t(T) =  \eta_F(t,T,F_t(T)) \,dW_t
}%
where the local volatility of futures prices is defined as
\Eq{
\eta_F(t,T,K) := \left( K - F_0(T) \left( 1 - e^{-\int_t^T a(u) \,du} \right) \right) \eta(t,k_F(t,T,K))
}%
\Eq{
k_F(t,T,K) := 1-\left(1-\frac{K}{F_0(T)}\right)e^{\int_t^T a(u) \,du}
}%
We can explicitly solve the above dynamics and we obtain
\Eq{
F_t(T) = F_0(T) \left( 1 - (1-s_t) e^{-\int_t^T a(u) \,du} \right)
}%

Then, we can calculate futures plain-vanilla options by means of an extended version of the Dupire equation. We define the normalized call price at time $0$ as given by
\Eq{
c_0(t,k) := \Ex{0}{\left(s_t - k\right)^{\!+}}
\,,\quad
c_0(0,k) = \left(1 - k\right)^{\!+}
}%
Option on futures can be expressed in term of normalized calls as
\Eq{
C_0(t,T,K) = P_0(T_p;e) \, F_0(T) \, e^{-\int_t^T a(u) \,du} c_0(t,k_F(t,T,K))
}%
where $P_0(T_p;e)$ is the price of a zero-coupon bond with yield $e_t$, with $e_t=0$ for futures margining style. In the following, we write $c(t,k)$ for $c_0(t,k)$ to ease the notation. By exploiting the linear form of the drift, we can derive the following parabolic PDE for normalized call prices.
\Eq{
\partial_t c(t,k) = \left( - a(t) - a(t) (1-k) \,\partial_k + \frac{1}{2} k^2 \eta^2(t,k) \,\partial^2_k \right) c(t,k)
}%
with the boundary conditions
\Eq{
c(t,0) = 1
\,,\quad
c(t,\infty) = 0
\,,\quad
c(0,k) = (1-k)^{\!+}
}%
The equation can be solved by means of the implicit method in a fast and efficient way as usually done for the standard Dupire equation.

In order to describe our calibration strategy, we look at the market. On energy markets we usually find the prices of plain-vanilla options (PVO) on futures contracts. Moreover, we can find also quotes of mid-curve options (MCO) and calendar spread options (CSO), even if with less liquidity. We recall that MCO are call and put options with maturity date occuring before the maturity of the underlying futures contract, while CSO are spread options between two subsequent futures. Both MCO and CSO contracts heavily depend on the time dependency of the instantaneous volatility of futures prices.

The model depends on two unknown deterministic functions: the mean reversion speed $a(t)$ and the local volatility $\eta(t,k)$. We choose a simple constant (time-independent) specification for the mean reversion, while we assume a non-parametric spline interpolation for the local volatility as in \cite{Nastasi2018}. We implement the following calibration procedure:
\begin{enumerate}
 \item we guess a value for the mean-reversion $a$,
 \item we perfectly calibrate the local volatility $\eta(t,k)$ to PVO prices,
 \item we evaluate all the MCO (or CSO) we wish to best fit,
 \item we repeat the procedure from the second step with a different value of $a$ if MCO (or CSO) prices are not recovered with the required precision.
\end{enumerate}

We test the calibration procedure on the TTF natural gas commodity market, since in the following sections we are interested in pricing swing options on this market. In particular we analyze futures contracts with delivery period of one month and options on such contracts. We compare the performance of the procedure against the fixed-point algorithm of \cite{Reghai2012}. We do not consider gradient-based optimization algorithms since they show poorer performances due to Jacobian evaluations (we have about one hundred parameters). The main improvements introduced in \cite{Nastasi2018} are using the accelerated fixed-point algorithm of \cite{Anderson1965}, and employing asymptotic expansions similar to \cite{Berestycki2002} to update the local volatility values.

\begin{figure}
\begin{center}
\scalebox{0.95}{%
\begin{tikzpicture}
\begin{semilogyaxis}[xlabel=Iterations,
                     ylabel=Calibration Error {[bp]},
                     ylabel style={overlay},
                     yminorticks=false,
                     grid=major,
                     legend style={legend pos=south west},
                     axis background/.style={fill=gray!10}]
  \addplot [color=blue,mark=o,mark options={style=solid},style=dashed] table [y=fe,x=N] from \calibngttf;
  \addplot [color=red,mark=triangle,mark options={style=solid},style=dashed] table [y=rbv,x=N] from \calibngttf;
  \addplot [color=blue,mark=o] table [y=AAfe,x=N] from \calibngttf;
  \addplot [color=red,mark=triangle] table [y=AArbv,x=N] from \calibngttf;
  \legend{asympt.,freez.,AA+asympt.,AA+freez.}
\end{semilogyaxis}
\end{tikzpicture}}
\caption{Calibration of 55 PVO on NG TTF Futures quoted on 29 March 2018 on ICE market with and without Anderson scheme (AA). Dashed red curve refers to \cite{Reghai2012} algorithm. Solid blue line to our implementation.}
\label{fig:calibngttf}
\end{center}
\end{figure}
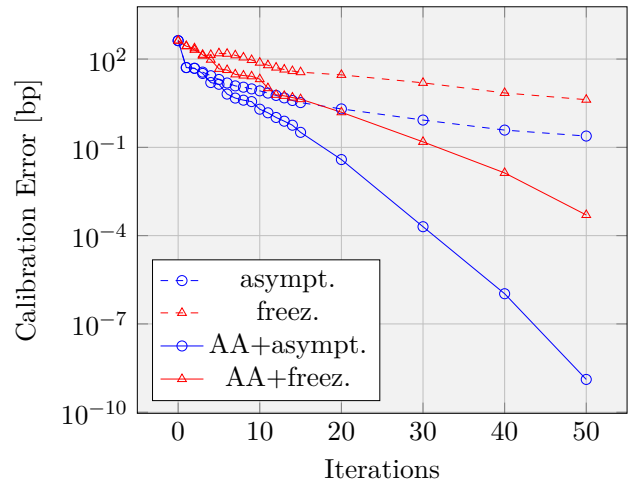

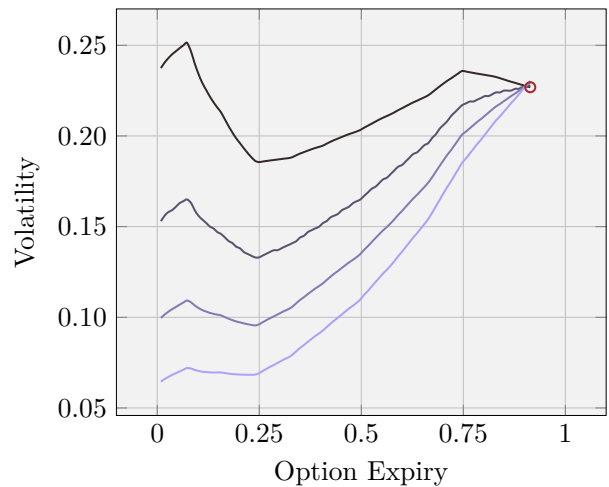
\begin{figure}
\begin{center}
\scalebox{0.95}{%
\begin{tikzpicture}
\begin{axis}[xlabel=Option Expiry,
                    y tick label style={/pgf/number format/.cd,fixed,fixed zerofill,precision=2,/tikz/.cd},
                    ylabel=Volatility,
                    ylabel style={overlay},
                    xmin=-0.1, xmax=1.1,
                    xtick={0.00,0.25,0.50,0.75,1.00},
                    grid=major,
                    axis background/.style={fill=gray!10}]
  \addplot [color=myblueA,smooth,thick] table [y=a0,x=T] from \mcottf;
  \addplot [color=myblueB,smooth,thick] table [y=a5,x=T] from \mcottf;
  \addplot [color=myblueC,smooth,thick] table [y=a10,x=T] from \mcottf;
  \addplot [color=myblueD,smooth,thick] table [y=a15,x=T] from \mcottf;
  \addplot [color=mydarkred,only marks,mark indices=99,mark=o,mark options={style=solid},thick] table [y=a0,x=T] from \mcottf;
\end{axis}
\end{tikzpicture}}
\caption{NG TTF One-Month MCO quoted on 29 March 2018 on ICE market. Volatilities quoted in the market (red dots) or implied by the model (blue lines). Mean reversion ranging from top to bottom from zero to $1.5$ in step of $0.5$.}
\label{fig:midcurvengttf}
\end{center}
\end{figure}

\begin{figure}
\begin{center}
\scalebox{0.95}{%
\begin{tikzpicture}
\begin{axis}[xlabel=Option Expiry,
                    ylabel=Volatility Drop {[bp]},
                    ylabel style={overlay},
                    xmin=-0.1, xmax=1.1,
                    xtick={0.00,0.25,0.50,0.75,1.00},
                    grid=major,
                    axis background/.style={fill=gray!10}]
  \addplot [color=myblueA,smooth,thick] table [y=a0,x=T] from \csottf;
  \addplot [color=myblueB,smooth,thick] table [y=a5,x=T] from \csottf;
  \addplot [color=myblueC,smooth,thick] table [y=a10,x=T] from \csottf;
  \addplot [color=myblueD,smooth,thick] table [y=a15,x=T] from \csottf;
\end{axis}
\end{tikzpicture}}
\caption{NG TTF One-Month CSO quoted on 29 March 2018 on ICE market. Volatility drops quoted in the market (red dots) or implied by the model (blue lines). Mean reversion ranging from top to bottom from $1.5$ to zero in step of $0.5$.}
\label{fig:calendarngttf}
\end{center}
\end{figure}
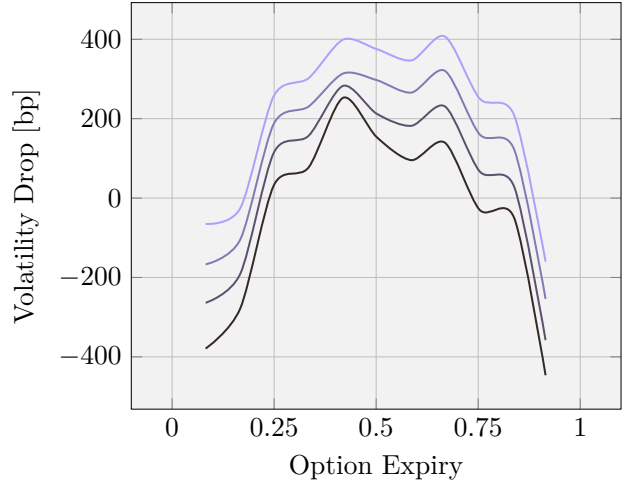

We show in Figure~\ref{fig:calibngttf} the performance of the calibration algorithm in determining the local-volatility function. We consider the market and model-implied volatility surfaces for quoted strikes and maturities, and we plot the maximum absolute difference between these two surfaces as a function of the algorithm iterations (we solve the Dupire equation for call prices on each iteration). We compare our results (blue solid line in figure) with alternative approaches, in particular the algorithm of \cite{Reghai2012} (red dashed line). We can see a that a maximum error of one tenth of basis point can be reached within twenty iterations. We refer to \cite{Nastasi2018} for complete a description of the algorithm and for other calibration examples. The example shown in the figure is obtained with mean reversion equal to zero, but we obtain similar results also with the other mean reversion values used in the present paper.

We continue by showing in Figure~\ref{fig:midcurvengttf} MCO for different choices of the mean-reversion parameter. We plot MCO implied volatilities as a function of the option expiry date. In the figure we can notice the impact of two effects arising when we change the mean-reversion parameter: on one hand (i) we are increasing futures volatility near its last trading date, on the other (ii) we calibrate PVO prices, so that the right-hand point in the figure remains stuck in the same place while the remaining of the curve moves downward as the the mean-reversion parameter increases in value.

Then, we represent in Figure~\ref{fig:calendarngttf} CSO for different choices of the mean-reversion parameter. We plot CSO volatility drops as a function of the option expiry date.  In a CSO contract on two consecutive futures we term volatility drop the difference between the implied volatility of the PVO on the second futures and the implied volatility of the MCO on the second futures with option expiry date on the last-trading date of the first future. The CSO plot shows the seasonality pattern usually found in volatilities quoted in the natural gas market. In the case of TTF natural gas we cannot find in the market liquid quotes, so that we proceed in the next sections by investigating different scenarios for the mean-reversion parameter. 

\subsection{Modelling Futures with Different Delivery Periods}
\label{sec:delivery}

We now continue the modelling section by extending the model presented in \cite{Nastasi2018} to deal with futures contracts with heterogeneous delivery periods.

Futures on commodities like natural gas, oil and electricity have as underlying a daily flow for the whole delivery period. Often day-ahead futures are quoted on the market as a close proxy of the spot prices. Moreover, futures on different delivery periods are usually quoted ranging from one day to a whole year. On the other hand, PVO contracts are usually quoted only for the most liquid delivery period.

A common way to model these futures prices consists in introducing a dynamics for futures with the shortest delivery period, and in building longer periods by summing futures prices. Yet, it is difficult to find models which allows to derive closed-form formulae for futures prices with longer periods. See for instance the approach of \cite{Benth2018}. Here, we rely on the linear form of the drift coefficient and on using a single risk factor to derive simple closed-form formulae for sums of futures prices.

We start by introducing the instantaneous futures price process $f_t(T)$ with delivery at time $T$, and we assume that we can model them by the local-volatility linear model presented in the previous section, so that we can write
\Eq{
f_t(T) = f_0(T) \left( 1 - (1-s_t) e^{-\int_t^T a(u) \,du} \right)
}
where the spot process is given by
\Eq{
ds_t = a(t) (1 - s_t) \,dt + \eta(t,s_t) s_t \,dW_t
\,,\quad
s_0 = 1
}%
Then, we calculate futures contracts with delivery periods $[T+\delta_0,T+\delta_1]$ as given by
\Eq{
F_t(T,\delta) = \int_0^\infty w(u-T,\delta) f_t(u) \,du
\,,\quad
w(\tau,\delta) := \frac{\ind{\delta_0 \le \tau \le \delta_1}}{\delta_1-\delta_0}
}%
where we define $\delta := \delta_1 - \delta_0$, and we discard the dependency on $\delta_0$ to lighten the notation. For instance, the futures contracts with a delivery period of one month presented in the previous section are now denoted as $F_t(T,{\rm 1m})$.

Now, we are left with the problem of deriving the dynamics of $F_t(T,\delta)$ for different delivery periods $\delta$. We can integrate the instantaneous futures over the delivery period
\Eq{
\label{eq:futures}
F_t(T,\delta) = F_0(T,\delta) \left(1 - (1-s_t(\delta)) e^{-\int_t^T A(u,\delta) \,du} \right)
}%
where we define
\Eq{
s_t(\delta) := 1 - (1-s_t) G(t,\delta)
\,,\quad
A(t,\delta) := a(t) - \partial_t \log G(t,\delta)
}%
\Eq{
G(t,\delta) := \frac{1}{F_0(t,\delta)} \int_0^\infty w(u-t,\delta) f_0(u) e^{-\int_t^u a(v) \,dv} \,du
}%

We notice that the relationship between $F_t(T,\delta)$ and $s_t(\delta)$ is the same holding between $f_t(T)$ and $s_t$ up to a change in parameters. In particular, we have $f_t(T) = F_t(T,0)$ and $s_t = s_t(0)$. We can calculate also the dynamics followed by the normalized spot price $s_t(\delta)$ corresponding to the delivery period $\delta$.
\Eq{
\label{eq:spot}
ds_t(\delta) = A(t,\delta) (1 - s_t(\delta)) \,dt + \eta(t,\delta,s_t(\delta)) s_t(\delta) \,dW_t
\,,\quad
s_0(\delta) = 1
}%
where we define the local volatility function
\Eq{
\eta(t,\delta,k) := \left(1-\frac{1-G(t,\delta)}{k}\right) \eta\!\left(t,1-\frac{1-k}{G(t,\delta)}\right)
}%
The following bounds are holding
\Eq{
s_t(\delta) > 1 - G(t,\delta)
\,,\quad
0 < G(t,\delta) \le 1
}%
so that the spot price is always positive.

The local-volatility linear model can be used when PVO on a single delivery period are actively quoted in market. We can calibrate the smile of the quoted delivery period, name it  $\bar\delta$, and we can imply the smile for other periods. For instance, in the natural gas market the only liquid options have as underlying asset futures contracts with a delivery period of one month. By a direct calculation we obtain a simple relationship linking the local volatilities of futures on different delivery periods. Indeed, we get
\Eq{
\eta(t,\delta,k)  = \left( 1 -  \frac{1}{k} \left( 1 - \frac{G(t,\delta) }{G(t,\bar\delta)}\right) \right) \eta\!\left( t , \bar\delta, 1 - (1-k) \frac{G(t,\bar\delta)}{G(t,\delta)} \right)
}%
which we evaluate for $k > 1 - G(t,\delta)$. The previous formula allows us to imply volatility smiles for any delivery period. Notice that smiles for different delivery periods can be different only if $a(t)>0$, since if $a(t)=0$ we get $G(t,\delta)=1$.

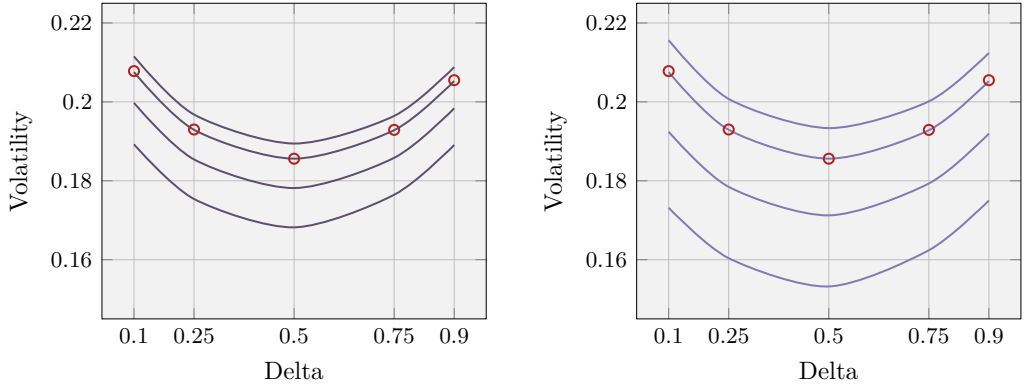
\begin{figure}
\begin{center}
\scalebox{0.95}{%
\begin{tikzpicture}
\begin{axis}[xlabel=Delta,
                    small,
                    width=0.55\textwidth,
                    ylabel=Volatility,
                    ylabel style={overlay},
                    xtick={0.1,0.25,0.5,0.75,0.9},
                    ymin=0.145,ymax=0.225,
                    grid=major,
                    axis background/.style={fill=gray!10}]
  \addplot [color=myblueB,smooth,thick] table [y=1m,x=d] from \vdttf;
  \addplot [color=myblueB,smooth,thick] table [y=1d05,x=d] from \vdttf;
  \addplot [color=myblueB,smooth,thick] table [y=3m05,x=d] from \vdttf;
  \addplot [color=myblueB,smooth,thick] table [y=6m05,x=d] from \vdttf;
  \addplot [color=mydarkred,only marks,mark=o,mark options={style=solid},thick] table [y=market,x=d] from \vdttf;
\end{axis}
\end{tikzpicture}
\hspace{1cm}
\begin{tikzpicture}
\begin{axis}[xlabel=Delta,
                    small,
                    width=0.55\textwidth,
                    ylabel=Volatility,
                    ylabel style={overlay},
                    xtick={0.1,0.25,0.5,0.75,0.9},
                    ymin=0.145,ymax=0.225,
                    grid=major,
                    axis background/.style={fill=gray!10}]
  \addplot [color=myblueC,smooth,thick] table [y=1m,x=d] from \vdttf;
  \addplot [color=myblueC,smooth,thick] table [y=1d10,x=d] from \vdttf;
  \addplot [color=myblueC,smooth,thick] table [y=3m10,x=d] from \vdttf;
  \addplot [color=myblueC,smooth,thick] table [y=6m10,x=d] from \vdttf;
  \addplot [color=mydarkred,only marks,mark=o,mark options={style=solid},thick] table [y=market,x=d] from \vdttf;
\end{axis}
\end{tikzpicture}}
\caption{NG TTF PVO on {\tt JUL18} futures quoted on 29 March 2018 on ICE market. Market (red dots) and model (blue lines) implied volatilities. Mean reversion equal to $0.5$ on left panel, to $1$ on right panel. Delivery periods ranging from top to bottom: 1 Day, One-Month, 3 Months, 6 Months.}
\label{fig:implvol}
\end{center}
\end{figure}

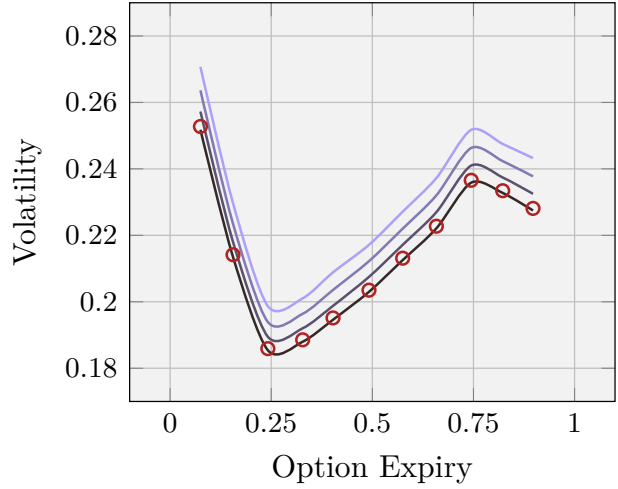
\begin{figure}
\begin{center}
\scalebox{1.2}{%
\begin{tikzpicture}
\begin{axis}[xlabel=Option Expiry,
                    small,
                    width=0.55\textwidth,
                    xmin=-0.1, xmax=1.1,
                    xtick={0.00,0.25,0.50,0.75,1.00},
                    ylabel=Volatility,
                    ylabel style={overlay},
                    ymin=0.17,ymax=0.29,
                    grid=major,
                    axis background/.style={fill=gray!10}]
  \addplot [color=myblueA,smooth,thick] table [y=DA_0,x=expiry_mkt] from \volada;
  \addplot [color=myblueB,smooth,thick] table [y=DA_05,x=expiry_mkt] from \volada;
  \addplot [color=myblueC,smooth,thick] table [y=DA_1,x=expiry_mkt] from \volada;
  \addplot [color=myblueD,smooth,thick] table [y=DA_15,x=expiry_mkt] from \volada;
  \addplot [color=mydarkred,only marks,mark=o,mark options={style=solid},thick] table [y=market,x=expiry_mkt] from \volada;
\end{axis}
\end{tikzpicture}}
\caption{NG TTF PVO quoted on 29 March 2018 on ICE market. Market (red dots) one-month futures PVO at-the-money volatilities. Model (blue lines) day-ahead futures PVO at-the-money implied volatilities. Mean reversion ranging from top to bottom from 1.5 to zero with a step of 0.5.}
\label{fig:implvolbyT}
\end{center}
\end{figure}

We show in Figure~\ref{fig:implvol} the volatility smiles implied by the model for different choices of the mean-reversion parameter in the case of the TTF natural gas market. The market is quoting the volatilities of PVO contracts on futures with one-month delivery period. The implied smiles maintain the same shape of the one-month smile because the market smile is almost symmetric in shape. In Figure~\ref{fig:implvolbyT} we plot the volatility backbones, namely the at-the-money implied volatilities as a function of option maturity.

We conclude this Section by hinting to a procedure to incorporate spikes in the day-ahead futures dynamics without altering the possibility to obtain closed-form formulae for futures prices, and to apply the calibration procedure to PVO, MCO and CSO described in the previous section. We do not consider spikes in the numerical parts. We leave this for a future work.

\subsection{Adding Spikes}
\label{sec:spike}

When looking at daily futures contracts, we could consider the impact of spikes in the day-ahead market. Here, we limit ourselves in describing a strategy to include spikes in the dynamics of the fictitious spot price by adapting the results of \cite{Hambly2009} to the local-volatility linear model. We leave to a future work the analysis of how calibrate spike parameters to observed spikes in the day-ahead market.

We can model spikes under the risk-neutral measure as a pure spike price process given by:
\Eq{
dy_t = -\gamma(t) y_t \,dt + \phi \,dN_t
\;,\quad
y_0 = 0
}%
where the spike mean-reversion speed $\gamma$ is a positive function, the amplitude $\phi$ is an exponentially distributed random variable with mean $\zeta$, $N_t$ is a Poisson process with intensity $\lambda(t)$ under the risk-neutral measure. We assume also that the fictitious spot and the spike process are independent. The process $y_t$ can be explicitly integrated leading to
\Eq{
y_t = \sum_{i=1}^{N_t} \phi_i \,\exp\left\{-\int_{\tau_i}^t \gamma(u) \,du\right\}
}%
where $\tau_i$ is the $i$-th jump time and $\phi_i$ is the corresponding amplitude realization. We can also calculate forward values in closed form as given by
\Eq{
\Ex{t}{y_T} = y_t e^{-\int_t^T \gamma(u) \,du} + h(t,T)
}%
where we define
\Eq{
h(t,T) := \zeta \int_t^T \lambda(u) e^{-\int_u^T \gamma(v) \,dv} \,du
}%

Adding spikes must leave unaltered the initial term structure of futures prices, so that we define spike-altered spot price as
\Eq{
{\bar s}_t := \frac{f_0(t)}{1+h(0,t)} \, (s_t + y_t)
\;,\quad
{\bar f}_t(T) := \Ex{t}{{\bar s}_T}
}%
where we assume that $s_t$ and $y_t$ are independent. We can proceed with the calculation of instantaneous futures prices.
\Eq{
{\bar f}_t(T) = f_0(T) \left( 1 - \frac{1-{\bar s}_t}{1+h(0,T)} e^{-\int_t^T a(u) \,du} - \frac{h(0,t)-y_t}{1+h(0,T)} e^{-\int_t^T \gamma(u) \,du} \right)
}%

Then, we integrate over the weights $w(\tau,\delta)$ to obtain the futures prices on longer delivery periods.
\Eq{
{\bar F}_t(T,\delta) = F_0(T,\delta) \left(1 - (1-s^h_t(\delta)) e^{-\int_t^T A^h(u,\delta) \,du} \right. \\ \left. - \,(H^{\gamma,h}(t,\delta)-y_t(\delta)) e^{-\int_t^T \Gamma^h(u,\delta) \,du} \right)
}%
where we define
\Eq{
s^h_t(\delta) := 1 - (1-s_t) G^{a,h}(t,\delta)
\;,\quad
y_t(\delta) := y_t G^{\gamma,h}(t,\delta)
}%
\Eq{
A^h(t,\delta) := a(t) - \partial_t \log G^{a,h}(t,\delta)
\;,\quad
\Gamma^h(t,\delta) := \gamma(t) - \partial_t \log G^{\gamma,h}(t,\delta)
}%
\Eq{
H^{\gamma,h}(t,\delta) := h(0,t) G^{\gamma,h}(t,\delta)
}%
in term of the deterministic functions
\Eq{
G^{a,h}(T,\delta) := \frac{1}{F_0(T)} \int_0^\infty\!\!\! w^h(u,T,\delta) f_0(u) e^{-\int_T^u a(v) \,dv} \,du
}%
\Eq{
G^{\gamma,h}(T,\delta) := \frac{1}{F_0(T)} \int_0^\infty\!\!\! w^h(u,T,\delta) f_0(u) e^{-\int_T^u \gamma(v) \,dv} \,du
}%
with modified weights
\Eq{
w^h(t,T,\delta) := \frac{w(T-t,\delta)}{1+h(0,t)}
}%

The calibration of plain-vanilla options on the longer-delivery futures prices $F_t(T,\delta)$ can be performed by mapping their prices onto the price of plain-vanilla options on the normalized spot process by conditioning on the process $y_t$, since it is independent of process $x_t$. The spike density $p_{y_t}$ can be calculated starting from the moment generating function of $y_t$. A closed-form solution in the case of time-homogeneous spike parameters and in the limit of high spike decay and small spike frequency can be found in \cite{Hambly2009}.

\section{Pricing Swing Options}
\label{sec:numerics}

We set up in this numerical section the stochastic control problem required to get swing option prices, and we solve it by means of a least-square Monte Carlo (LSMC) simulation. As a specific example we consider swing options traded in the TTF natural gas market.

The LSMC algorithm is particularly effective when we have to deal only with few risk factors, since the method requires to calculate a linear regression whose dimension rapidly explodes as the number of risky factors increases. In our case we adopt a parsimonious model with only one risk factor. However, for a better description of curve and smile dynamics we could look at model extensions inclusive of additional risk factors as described in \cite{Nastasi2018}. For this reason we will also investigate in Section~\ref{sec:rl} solutions which could be applied in higher dimensionality settings.

\subsection{Contract Description}

A swing option contract guarantees a flexible daily supply of gas with a delivery period of one month. The underlying contracts are the day-ahead futures, namely $F_{T_i}(T_{i+1},1{\rm d})$ for each fixing date $T_1,\ldots,T_{n_f}$ within the delivery period. At each fixing date the owner of the option is allowed to buy a quantity $N_{T_i}$ of gas within a daily range $[N_m,N_M]$ at a strike price $K$. The total consumption of gas must be within a total range $[C_m,C_M]$. The option price can be written as
\Eq{
W_0 := \max_{N\in{\cal N}} \sum_{i=1}^{n_f} \,\Ex{0}{ N_{T_i} ( F_{T_i}(T_{i+1},1{\rm d}) - K ) } P_0(T_{p,i};e)
}%
where $P_0(T_{p,i};e)$ is the price of a zero-coupon bond with yield $e_t$, and the consumption plan $N := \{N_{T_1},\ldots,N_{T_{n_f}}\}$ can be chosen from a set $\cal N$ of plans subject to the following constraints.
\Eq{
N_m \leq N_{T_i} \leq N_M
\;,\quad
C_m \leq \sum_{i=1}^{n_f} N_{T_i} \leq C_M
}

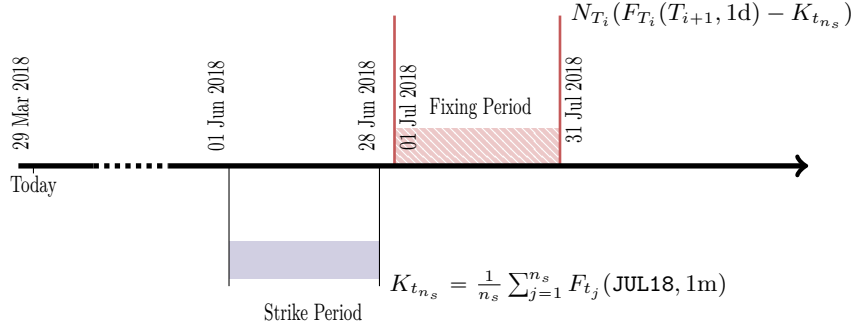
\begin{figure}
\begin{center}
\begin{tikzpicture}

\fill [mydarkred!25] (5,0) rectangle (7.2,0.5) node[above,color=black,xshift=-30pt]{{\footnotesize \scalebox{.7}[1.0]{Fixing Period}}};
\fill [pattern=north west lines,pattern color=white] (5,0) rectangle (7.2,0.5);
\fill [mydarkblue!25] (2.8,-1.5) rectangle (4.8,-1.0) node[below,xshift=-25pt,yshift=-20pt,color=black] {{\footnotesize \scalebox{.7}[1.0]{Strike Period}}};

\draw[line width=1pt,color=mydarkred!75] (7.2,0) -- (7.2,2) node [right,color=black] {{\footnotesize$N_{T_i} ( F_{T_i}(T_{i+1},1{\rm d}) - K_{t_{n_s}} )$}};
\draw[line width=1pt,color=mydarkred!75] (5,0) node [right,color=black,rotate=90,xshift=1.9pt,yshift=-5pt] {{\footnotesize \scalebox{.7}[1.0]{01 Jul 2018}}} -- (5,2);

\draw[line width=2pt] (0,0) -- (1,0);
\draw[dotted,line width=2pt] (1,0) -- (2,0);
\draw[->,line width=2pt] (2,0) -- (10.5,0);

\draw (0.2,0) node [below] {{\footnotesize \scalebox{.7}[1.0]{Today}}} -- (0.2,-0.1) node [right,rotate=90,xshift=5pt,yshift=5pt] {{\footnotesize \scalebox{.7}[1.0]{29 Mar 2018}}};
\draw (7.2,0) -- (7.2,0.0) node [right,rotate=90,xshift=5pt,yshift=-5pt] {{\footnotesize \scalebox{.7}[1.0]{31 Jul 2018}}};

\draw (2.8,0) node [right,rotate=90,xshift=1.9pt,yshift=5pt] {{\footnotesize \scalebox{.7}[1.0]{01 Jun 2018}}} -- (2.8,-1.6);
\draw (4.8,0) node [right,rotate=90,xshift=1.9pt,yshift=5pt] {{\footnotesize \scalebox{.7}[1.0]{28 Jun 2018}}} -- (4.8,-1.6) node [right,] {{\footnotesize$K_{t_{n_s}} = \frac{1}{n_s} \sum_{j=1}^{n_s} F_{t_j}(\texttt{JUL18},{\rm 1m})$}};

\end{tikzpicture}
\caption{Term-sheet data for a swing option contract with delivery in July 2018 in TTF natural gas market. The strike is fixed by avering the {\tt JUL18} futures contract observed in the month of June.}
\label{fig:swing}
\end{center}
\end{figure}

The strike price of swing options can be known at inception, or fixed at a forward date. In the latter case it is calculated as the daily average of a specific one-month futures contract over the observation dates $t_1,\ldots,t_{n_s}$. For example, the strike price of a swing option with delivery in July 2018 is fixed by averaging the daily observations of the {\tt JUL18} one-month contract, namely we set
\begin{equation*}
K_{t_{n_s}} := \frac{1}{n_s} \sum_{j=1}^{n_s} F_{t_j}(\texttt{JUL18},{\rm 1m})
\end{equation*}%
where $t_1$ is the 1st of June 2018 and $t_{n_s}$ is the 28th of the same month (last trading date).

We show in Figure~\ref{fig:swing} the {\tt JUL18} swing option on NG TTF day-ahead futures term-sheet data. In this example the strike price is set by observing one-month futures contracts.

\subsection{Least-Square Monte Carlo Simulation}

We can write the stochastic control problem underlying the pricing of a swing option contract by introducing the consumption strategy $N_{T_i}$ which represents the quantity of gas delivered in $T_i$, and by defining the total consumption up to time $T_i$ as given by
\Eq{
C_{T_i} := \sum_{j=1}^i N_{T_j}
}%
Thus, on each day $T_i$ in the simulation we have to solve the following control problem
\Eq{
W_{T_i} = \max_{N_{T_i}} \left\{ N_{T_i}( F_{T_i}(T_{i+1},{\rm 1d}) - K ) + \ExCo{ W_{T_{i+1}}(N_{T_i}) \frac{P_0(T_{p,i+1};e)}{P_0(T_{p,i};e)} }{F_{T_i},C_{T_{i-1}}} \right\}
 \label{eq: control problem}
}%
In case the option is forward starting we should add to conditioning factors also the strike price.

We can solve the control problem by means of a LSMC simulation. Here, we describe the details of our implementation, which can be split into three steps: (i) we build consumption grids, (ii) we estimate of the value functions on the grids by means of regressions with a backward procedure, (iii) we compute the swing option price with a standard forward Monte Carlo simulation.

 For simplicity, the method is described in the following by considering zero interest rates and fixed strike prices.

\subsubsection{Building Consumption Grids}

We start by constructing the consumption grid by taking care that the points corresponding to extreme choices of the amount to be consumed are included. We define the global constraint functions at each fixing date $T_i$ as given by
\Eq{
U_i := \min \left( \left(C_M - N_m\right) (n_f - i) \,,\; i N_M \right)
}%
and
\Eq{
D_i := \max \left( \left(C_m - N_M\right) (n_f - i) \,,\; i N_m \right)
}%
At each date $T_i$ the total consumption must be within such values: $D_i \le C_{T_i} \le U_i$. We define $C^{i}$ as the vector representing the consumption grid at fixing date $T_i$. The consumption grid is built by following Algorithm~\ref{algo: grid}.

\begin{algorithm}
  \begin{algorithmic}[1]
    \Procedure{Grid}{$\{ T_i \}_{i=1}^{n_f},N_m, N_M, C_m, C_M, \Delta$}
      \State $C^{0} := [0]$\Comment{Starting consumption}
      \For{$i=1$ \texttt{to} $n_f$}
        \For{$x$ \texttt{in} $C^{i-1}$}\Comment{Bang Bang points}
            \State \texttt{append} $\min \left( U_i \; , \; x + N_M \right)$ \texttt{to} $C^{i}$
            \State \texttt{append} $\max \left( D_i \; , \; x + N_m \right)$ \texttt{to} $C^{i}$
        \EndFor
        \State $C^{i}$ = \texttt{unique}($C^{i}$)\Comment Sort the grid and erase duplicates
        \For{$j$ \texttt{in} $\texttt{length}(C^{i}$) - 1}\label{algo: grid: last_for_begin}
            \State \texttt{append} \texttt{unif}$(C^{i}_j, C^{i}_{j + 1}, \Delta)$ \texttt{to} $C^{i}$\Comment{Thicken the grid}
        \EndFor \label{algo: grid: last_for_end}
        \State $C^{i}$ = \texttt{unique}($C^{i}$) \Comment Sort the grid and erase duplicates
      \EndFor
    \EndProcedure
  \end{algorithmic}
  \caption{Algorithm to build the consumption grid. Operations from \ref{algo: grid: last_for_begin} to \ref{algo: grid: last_for_end} are performed only when the strategy is continuous.}
  \label{algo: grid}
\end{algorithm}

\begin{figure}
 \begin{center}
  \includegraphics[scale=0.4]{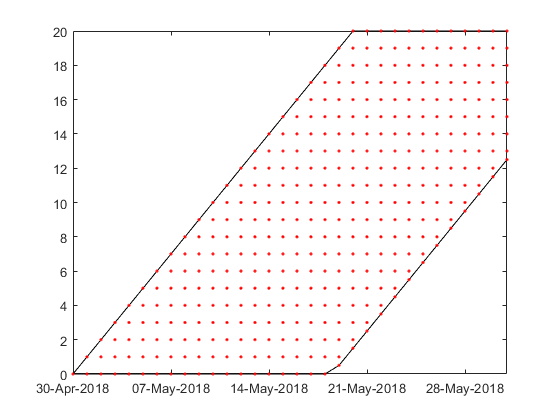}
  \includegraphics[scale=0.4]{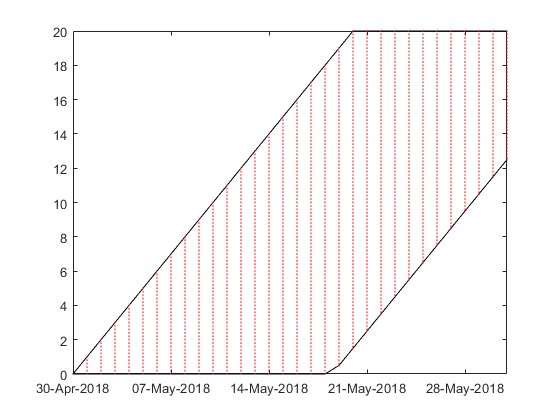}
 \end{center}
 \caption{Grids obtained for a swing delivering in the month of April 2019 with $N_M=1$, $N_m=0$, $C_M=15.7$, $C_m=5.2$ and $\Delta=\frac{1}{6}$. Left and right panel show the bang bang and continuous cases respectively.}
 \label{fig: grid}
\end{figure}

 On each date $T_i$ we start by adding the points allowed by a bang-bang strategy. We use the term bang-bang as in \cite{Jaillet2004} to indicate a strategy which on each date $T_i$ is consuming the minimum or the maximum amount of commodity according to all the constraints. Thus, defining the starting consumption $C^0$ as a vector with a single component equal to zero, at time $T_1$ we have only two possible bang-bang states given by $N_m$ and $N_M$ (but in the case of tighter global constraints). On the following date $T_2$ the grid has three bang-bang points, obtained by starting from the consumption levels of the previous date and consuming the minimum or maximum allowed quantities, and so on. Then, we refine the grid in between the bang-bang points to allow for intermediate choices (continuous consumption strategy). The resulting grids are depicted in Figure~\ref{fig: grid}.

\subsubsection{Calculating the Regression Coefficients}

Now, we proceed by describing the simulation algorithm. We start by producing a Monte Carlo simulation on dates $T_i$ for the day-ahead futures prices according to Equations~\eqref{eq:futures} and~\eqref{eq:spot}, and   we build the consumption grids $C^{i}$ according to the previous algorithm. We call $F^{(k)}_{T_i}$ the $i-$th fixing of the $k-$th simulation. For each point $C^{i - 1} _\ell$ of the grid at previous time, we introduce the set $N(C^{i - 1}_\ell)$ of all the possible consumption levels, whose $j$-th element can be defined as
\Eq{
 N_j(C^{i - 1}_\ell) := C^{i}_j - C^{i - 1}_\ell
}%
and the the set $\mathcal{Q}^{T_i}(C^{i - 1}_\ell)$ of admissible consumption levels given global and local constraints relative to the $\ell-$th point of the grid
\Eq{
 \mathcal{Q}^{T_i}(C^{i - 1}_\ell) := \left\{N \in N(C^{i - 1}_\ell) \bigcap \left[ N_m, N_M \right] \, | \, C^{i - 1}_\ell + N \in \left[ C_m, C_M \right] \right\}
}%
Then, we can write the control problem on the grid as given by
\Eq{
 W_{T_i} \left(F^{(k)}_{T_i}, C^{i - 1}_\ell \right) = \max_{N \in \mathcal{Q}^{T_i}(C^{i-1}_\ell)} \left\{ N \left(F^{(k)}_{T_i} - K \right) + \ExCo{ W_{T_{i + 1}} \left( F_{T_{i}}, C^{i - 1}_\ell + N \right) }{F_{T_i} = F^{(k)}_{T_i}} \right\}
 \label{eq: discrete problem}
}%
with terminal condition
\Eq{
 W_{T_{n_f}} \left(F^{(k)}_{T_{n_f}}, C^{n_f - 1}_j \right) =
 \begin{cases}
  \min \left(U_{n_f} - C^{n_f - 1}_j \; , \; N_M \right) \left( F_{T_{n_f}}^{(k)} - K \right) \qquad F_{T_{n_f}}^{(k)} > K \\
  \max \left(C^{n_f - 1}_j - D_{n_f} \; , \; N_m \right) \left( F_{T_{n_f}}^{(k)} - K \right) \qquad F_{T_{n_f}}^{(k)} \leq K
 \end{cases}
}%

We can solve the problem backward in time by starting from the terminal condition in $T_ {n_f}$, and proceeding to the previous steps by numerically evaluating the forward expectation in the right-hand side of Equation~\eqref{eq: discrete problem} by means of the Monte Carlo simulation. We call $f_{T_i} \left(F; C^{i - 1}_\ell + N \right)$ the estimate of such forward expectation
\Eq{
\ExCo{ W_{T_{i + 1}} \left( F_{T_{i + 1}}, C^{i - 1}_\ell + N \right) }{F_{T_i} = F^{(k)}_{T_i}} \approx f_{T_i} \left(F; C^{i - 1}_\ell + N \right)
}%
and we suppose that it is quadratic with respect to day-ahead futures prices:
\Eq{
\label{eq: value function estimation}
f_{T_i} \left(F; C\right) := \alpha^{i}\left(C\right) + \beta^{i}\left(C\right) F + \gamma^{i}\left(C\right) F^2
}%
We note that $C^{i - 1}_\ell + N \in C^{i}$ by construction, hence we can estimate the coefficients by regressing for each $j$ the realizations
\Eq{
y^{(k)} := W_{T_{i + 1}} \left( F^{(k)}_{T_{i + 1}}, C^{i}_j \right)
}%
against
\Eq{
x^{(k)} := F^{(k)}_{T_{i}}
}%
The problem given by Equation~\eqref{eq: discrete problem} at each time $T_i$ before the terminal condition is then solved by replacing the estimate just performed \eqref{eq: value function estimation} in place of the forward expectation. The procedure just described is repeated backward in time until the first fixing.

\subsubsection{Pricing with a Forward Simulation}

Once the backward procedure is completed we have calculated the coefficients $\alpha^{i}$, $\beta^{i}$ and $\gamma^{i}$ on each grid date $T_i$ which allows us to approximate the forward expectation given by Equation~\eqref{eq: value function estimation} on any scenario. Thus, in order to avoid biases, we proceed by sampling a second Monte Carlo simulation for day-ahead futures prices. On each scenario $k$ of the second simulation and on each date $T_i$ we calculate $F^{(k)}_{T_i}$. Starting from the first fixing, at each step, being at a certain point $C_{i - 1}^{(k)}$ on the grid $C^{i - 1}$, we choose the quantity to consume $\hat{N}^{(k)}_{i}$ solving the problem optimization \eqref{eq: discrete problem} with the coefficients calculated in the previous simulation. This step takes us to the point $C_{i}^{(k)} = C_{i - 1}^{(k)} + \hat{N}^{(k)}_{i}$. Repeating the step described until reaching the last fixing we get the reward
\begin{equation}
 R^{(k)}_{T_{n_f}} := \sum_{i=1}^{n_f} \hat{N}^{(k)}_{i} \left( F^{(k)}_{T_i} - K \right)
\end{equation}
Hence, the swing option price is given by averaging the rewards.
\begin{equation}
W_{T_0} \left(F_{T_0}; 0 \right) = \Ex{T_0}{R_{T_{n_f}}}
\end{equation}

\subsection{Numerical Investigations with LSMC}
\label{sec:lsmc}

We are now ready to calculate the price of swing options with the local-volatility linear model by using the LSMC algorithm. It is our aim to highlight the impact of the mean-reversion parameter in swing option prices. Moreover, we wish to show that Theorem 2 in \cite{Bardou2009} is holding, and the LSMC algorithm is able to select the optimal strategies in agreement with the theorem.

We consider for our numerical analysis futures contracts on the TTF natural gas, quotations are expressed in \euro/MWh. We calibrate our model to PVO quoted on 29 March 2018 on ICE market. We consider swing option contracts with delivery ranging from May 2018 up to June 2019. We consider both fixed-strike option and floating-strike options with at-the-money strike. The strike price is calculated by considering the one-month futures contract delivering in the same period of delivery of the swing contract. Fixed-strike options evaluate the futures price at contract inception, while floating-strike options make a daily average of the futures prices on a time window starting after contract inception and ending before delivering, as previously depicted in Figure~\ref{fig:swing}. All contracts, if not specified otherwise, have daily and global constraints given by
\Eq{
N_m = 0 \,{\rm MWh}
\;,\quad
N_M = 1 \,{\rm MWh}
}
\Eq{
C_m = 12.5 \,{\rm MWh}
\;,\quad
C_M = 20 \,{\rm MWh}
}%
Notice that we choose the daily constraints without loss of generality by following what is usually done in the literature, see \cite{Bardou2009}. Different choices can be obtained by simply scaling all the relevant quantities.

\subsubsection{Fixed vs.\ Floating-Strike Options}

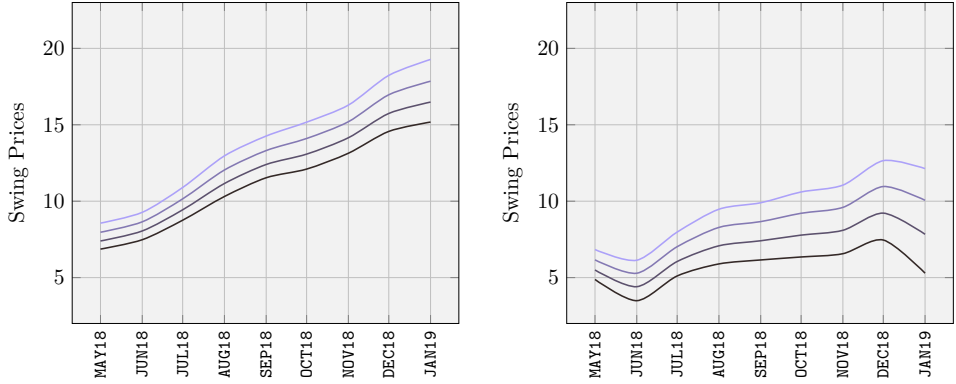
\begin{figure}
\begin{center}

\scalebox{0.75}{%
\begin{tikzpicture}
\begin{axis}[
                    ylabel=Swing Prices,
                    ylabel style={overlay},
                    ymin=2, ymax=23,
                    xmin=4.3200, xmax=4.3487,
                    xtick=data,
                    xticklabels={
                                 \footnotesize \rotatebox{90}{\tt MAY18},
                                 \footnotesize \rotatebox{90}{\tt JUN18},
                                 \footnotesize \rotatebox{90}{\tt JUL18},
                                 \footnotesize \rotatebox{90}{\tt AUG18},
                                 \footnotesize \rotatebox{90}{\tt SEP18},
                                 \footnotesize \rotatebox{90}{\tt OCT18},
                                 \footnotesize \rotatebox{90}{\tt NOV18},
                                 \footnotesize \rotatebox{90}{\tt DEC18},
                                 \footnotesize \rotatebox{90}{\tt JAN19}},
                    grid=major,
                    axis background/.style={fill=gray!10}]
  \addplot [color=myblueA,smooth,thick] table [y=mr0,x=fixing_start] from \swingkfix;
  \addplot [color=myblueB,smooth,thick] table [y=mr05,x=fixing_start] from \swingkfix;
  \addplot [color=myblueC,smooth,thick] table [y=mr1,x=fixing_start] from \swingkfix;
  \addplot [color=myblueD,smooth,thick] table [y=mr15,x=fixing_start] from \swingkfix;
\end{axis}
\end{tikzpicture}
\hspace*{1cm}
\begin{tikzpicture}
\begin{axis}[
                    ylabel=Swing Prices,
                    ylabel style={overlay},
                    ymin=2, ymax=23,
                    xmin=4.3200, xmax=4.3487,
                    xtick=data,
                    xticklabels={
                                 \footnotesize \rotatebox{90}{\tt MAY18},
                                 \footnotesize \rotatebox{90}{\tt JUN18},
                                 \footnotesize \rotatebox{90}{\tt JUL18},
                                 \footnotesize \rotatebox{90}{\tt AUG18},
                                 \footnotesize \rotatebox{90}{\tt SEP18},
                                 \footnotesize \rotatebox{90}{\tt OCT18},
                                 \footnotesize \rotatebox{90}{\tt NOV18},
                                 \footnotesize \rotatebox{90}{\tt DEC18},
                                 \footnotesize \rotatebox{90}{\tt JAN19}},
                    grid=major,
                    axis background/.style={fill=gray!10}]
  \addplot [color=myblueA,smooth,thick] table [y=mr0,x=fixing_start] from \swingkfwd;
  \addplot [color=myblueB,smooth,thick] table [y=mr05,x=fixing_start] from \swingkfwd;
  \addplot [color=myblueC,smooth,thick] table [y=mr1,x=fixing_start] from \swingkfwd;
  \addplot [color=myblueD,smooth,thick] table [y=mr15,x=fixing_start] from \swingkfwd;
\end{axis}
\end{tikzpicture}}

\end{center}
\caption{Swing option prices by varying the delivery starting date. Mean reversion ranging from top to bottom from 1.5 to 0 with a step of 0.5. Left panel fixed-strike contracts. Right panel floating-strike contracts.}
\label{fig:fixfloat}
\end{figure}

We start by analyzing the impact of the mean-reversion speed on swing option prices. In Figure~\ref{fig:fixfloat} we show the swing option prices for different delivery periods, each period corresponds to the delivery of futures contract quoted in the market. The trend of the price of the fixed strike options  with respect to the delivery month can be easily understood. As time increases, the option increases its time value which translates into an increase in price. The increasing trend as a function of the mean reversion speed is instead explained by the two graphs in Figure~\ref{fig:implvol}. This picture shows that the volatility of the day-ahead contract, i.e.\ one-day delivery period, implied by the model, is increasing as the mean reversion speed increases, which translates into an increase in the price of the swing option. On the contrary, if we look at the prices of the forward start options for mean reversion speed equal to 0, we note that the price trend reproduces the shape of the at the money market volatility at Figure~\ref{fig:implvolbyT}. This is due to the fact that the volatility of the monthly Futures observed during the strike period is equal to the volatility of the day-ahead contract. By increasing the mean reversion, instead, we have the two opposite effects: the volatility of the strike decreases as shown in the Figure~\ref{fig:midcurvengttf} while the volatility of the day-ahead contract, as already said, increases. As result the forward volatility relative to the fixing period increases, producing increasing prices as the mean reversion speed increases.

In the following sections we will focus on specific numerical problems, so that we will consider only the case of a fixed-strike option delivering in May 2018. Moreover, we set the mean reversion speed to $1$.

\subsubsection{Bang-Bang Strategies}\label{sec: bang-bang}

We continue by investigating the strategies selected by the LSMC algorithm. We recall that Theorem 2 in \cite{Bardou2009} describes the structure of optimal strategies for swing options when the consumption levels have a specific form. In particular, it states that the if the minimum global constraint and the difference between the global constraints can be expressed as an integer multiple of the difference between the daily constraints, then the optimal strategy on all dates is consuming the daily minimum or maximum (that strategy is of bang-bang type). For instance, the swing option contract used for the example of Figure~\ref{fig:fixfloat} does not satisfy the theorem, while the same contract with integer values for the global constraints is within the theorem since the difference between the local constraints is $1$.

We start from one of the cases studied in the previous section (fixed-strike delivering in May 2018 with $a=1$). Such case does not satisfy the hypotheses of the Theorem since the minimum global constraint is not an integer number. The price of the corresponding swing option contract can be read in the top-left entry of Table~\ref{tab:bb}. Then, we consider three different scenarios.
\begin{enumerate}
 \item We limit the allowed strategies to be only of bang-bang type. We expect to see a reduced price in this case since we forbid intermediate consumption choices. Indeed, in the top-right entry of the table we obtain a lower price.
 \item Without limitations on the strategies we change the minimum global constraint to $12\,{\rm MWh}$, so that now we satisfy the hypotheses of the Theorem. In the bottom-left entry of the table we report the price of this scenario. The price is now bigger since the global constraints are wider.
 \item With the constraint of the previous scenario we limit the allowed strategies to be only of bang-bang type. We expect not to see a reduced price in this case since we forbid consumption choices which are not selected for the optimal strategy. Indeed, in the bottom-right entry of the table we can see that the price is unchanged.
\end{enumerate}

\begin{table}
\begin{center}
\begin{tabular}{|r|c c|}
    \hline
                                                & {All Strategies}    &  {Only Bang-Bang} \\ \hline
    {Out of Theorem Hyp.} & $7.97 \pm 0.04$ & $7.76 \pm 0.04$ \\ \hline
    {Within Theorem Hyp.} & $8.46 \pm 0.04$ & $8.46 \pm 0.04$ \\
    \hline
\end{tabular}
\end{center}
\caption{Swing option prices in four different cases. We consider the reference case with $C_m=12.5\,{\rm MWh}$ not satisfying the Theorem 2 in \cite{Bardou2009}, and a variant by changing the constraint to $C_m=12\,{\rm MWh}$ so that the Theorem is satisfied. Prices are calculated either allowing all possible strategies or only the bang-bang ones. One-sigma statistical errors are displayed.}
\label{tab:bb}
\end{table}

\begin{figure}
\begin{center}
\includegraphics[scale = 0.42]{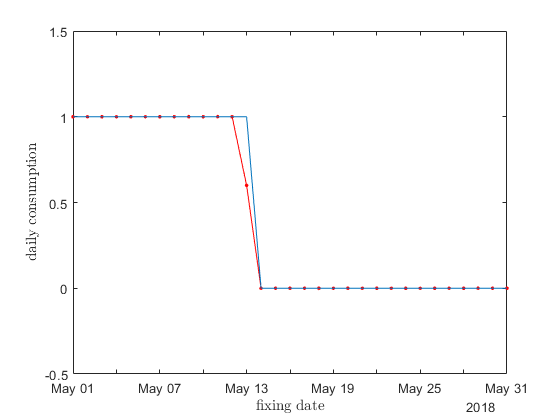}
\includegraphics[scale = 0.42]{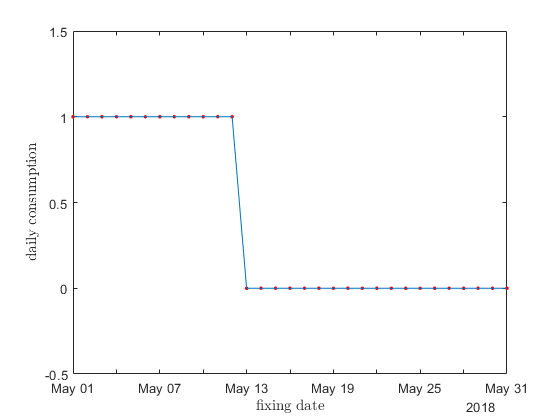}
\caption{Daily consumption selected by the optimal strategy for a fixed-strike swing option delivering in May 2018. Left panel displays the out-of-theorem scenario, while the right panel the within-theorem. Mean reversion speed $a=1$. The blue lines refer to the case where only bang-bang strategies are allowed, while red lines to the case without restrictions. In the bang-bang case (right panel) the two lines coincide.}
\label{fig:consumption}
\end{center}
\end{figure}

We support our discussion by showing in Figure~\ref{fig:consumption} the graph of daily consumption $N_{T_i}$ selected by the optimal strategy on a particular simulation path, when we assume that the minimum global constraint is either $12.5\,{\rm MWh}$ (out-of-theorem hypotheses, left panel) or $12\,{\rm MWh}$ (within-theorem hypotheses, right panel). When we are out of the Theorem hypotheses we can see that exists an optimal strategy (red line) which is not of bang-bang type, and we are able to exploit these strategies.

\section{Optimal Strategies via Reinforcement Learning}
\label{sec:rl}

As we have seen swing option pricing requires to solve a stochastic control problem with a continuous set of actions. Standard techniques rely on regression-based simulations whose performances may degrade when the dimensionality of the problem increases. In particular, if we wish to extend our analysis to long-dated options, we should introduce more driving factors to deal with the curve dynamics and possibly of the volatility dynamics. Indeed, in \cite{Nastasi2018} we extend the local-volatility linear model in this direction.

In the literature different techniques are investigated starting from the results of \cite{Barrera2006} on the form of the optimal consumption strategy. These authors prove that in the case of differentiable constraints the optimal strategy has the so-called bang-bang form, namely on each date the optimal strategy is delivering the minimum or the maximum allowed by the constraints. In \cite{Bardou2009} such result is extended also to sharp constraints when the contract specifics have very particular forms. For contracts with a bang-bang optimal strategy it is possible to simplify the stochastic control problem since we have only two choices at each date, leading to a simpler LSMC algorithm.

Here, we wish to price swing option with an alternative algorithm based on reinforcement learning (RL) techniques. RL has been introduced in finance to assist the trading activity. See for instance \cite{Kolm2019}. See also \cite{Becker2019} for applications to American options. In \cite{Barrera2006} RL is already considered as a possible pricing tool for swing options. In our approach we use the recently developed proximal policy optimization (PPO) algorithm proposed in \cite{Schulman2017}.

\subsection{Proximal Policy Optimization Algorithm}
\label{sec:ppo}

RL describes how an agent behaves in an environment so to maximize some notion of cumulative reward. The actions of the agent as a function of his observations of the environment are termed the agent policy. In our case the agent can choose the amount of commodity to be delivered within the contract limits, so that the policy is the consumption strategy, while the rewards are the cash flows generated by holding the swing option. Once the agent is trained, and the optimal policy is selected, we can run a Monte Carlo simulation to calculate the swing option price.

\subsubsection{Agent Interaction with the Environment}

\begin{figure}
\begin{center}
\begin{tikzpicture}[node distance = 9em, auto, thick]
  \node [bluebox] (A1) {\shortstack{Agent\\[0.5ex] {\small $\{N^\theta_{T_i},V^\theta_{T_i}\}$}}};
  \node [yellowbox, above left of=A1] (S1) {\shortstack{State\\[0.5ex] {\small $\{T_i,F_{T_i},C^\theta_{T_{i-1}}\}$}}};
  \node [fadedyellowbox, above right of=A1] (S2) {\shortstack{State\\[0.5ex] {\small $\{T_{i+1},F_{T_{i+1}},C^\theta_{T_i}\}$}}};
  \node [redbox, below left of=A1] (R1) {\shortstack{Reward\\[0.5ex] {\small $N^\theta_{T_i} (F_{T_i}-K)$}}};
  \node [fadedredbox, below right of=A1] (R2) {\shortstack{Reward\\[0.5ex] {\small $N^\theta_{T_{i+1}} (F_{T_{i+1}}-K)$}}};
  
  \draw [fixto] (S1) -- node [below, midway, xshift=-5ex, yshift=2ex] {\shortstack{read\\environment}} (A1);
  \draw [fixto] (R1) -- node [above, midway, xshift=-4ex, yshift=-1ex] {\shortstack{get\\reward}} (A1);
  \draw [fixto, draw=gray] (S2) -- (A1);
  \draw [fixto, draw=gray] (R2) -- (A1);
  \draw [fixto, bend right] (A1.0) to node [right, near start, yshift=-0.25ex] {\shortstack{update\\consumption}} (S2);
  \draw [floatto, draw=mydarkyellow!75!black!100] (S1) -- node [above, midway, yshift=0.5ex] {simulate} (S2);
  \draw [floatto, draw=mydarkred!75!black!100] (R1) -- node [below, midway, yshift=-0.5ex] {cumulate} (R2);
\end{tikzpicture}
\caption{Agent description.}
\label{fig:agent}
\end{center}
\end{figure}
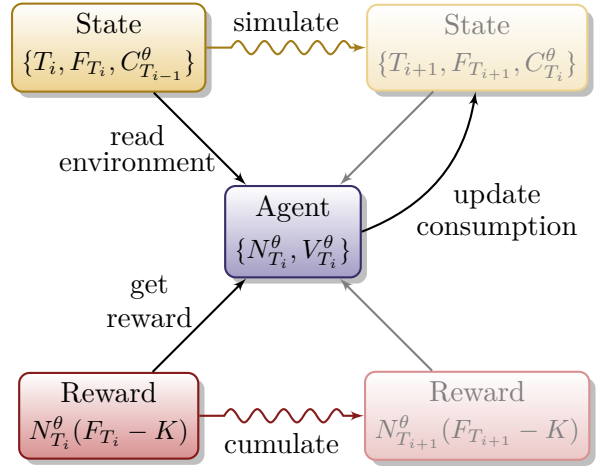

We consider as before a discrete time-grid of fixing times $T_1,\ldots,T_{n^f}$. The algorithm we chose for the training of the agent belongs to the family of actor-critic algorithms. In particular, in our setting, this means that the agent uses a parametric function with parameters $\theta$ to calculate both the quantity $N^\theta_{T_i}$ to consume at time $T_i$, and its best estimate of the value function $V^\theta_{T_i}$; the latter represents the expected value of future rewards, which will match the option price $W_{T_i}$ for optimal $N^\theta$. The agent makes its decision by observing the environment given by the fixing time $T_i$, the day-ahead futures price $F_{T_i} := F_{T_i}(T_{i+1},{\rm 1d})$, and the total quantity of gas $C^\theta_{T_{i-1}}$ consumed up to time $T_{i-1}$. We represent in Figure~\ref{fig:agent} the relationships between the agent and the environment. 

We adopt as learning strategy the PPO algorithm developed in \cite{Schulman2017}. This algorithm is well-suited for continuous control problems\footnote{We use the implementation of the algorithm found in OpenAI Baselines \url{https://github.com/openai/baselines}}. The PPO algorithm collects a small batch of experiences interacting with the environment to update its decision-making policy. The expected reward and the value function of a new policy are estimated by sampling from the environment.  A brief overview of how this is done is provided here below.

\subsubsection{Description of the Learning Strategy}

In the PPO algorithm the policies are randomized, so they are defined as probability distributions on the set of possible actions. In our case they represent the probability of a specific gas consumption on each fixing date given the value of the day-ahead futures contract and the total level of gas consumption up to the previous fixing date. In case of forward-strike swing options we have to include also the strike level. If the space of controls is a continuum, the algorithm considers the actions to be random variables $\tilde{N}^{\theta}_{T_i}$ distributed according to independent Gaussian distributions $\phi$ centered on the value $N^\theta_{T_i}$, which is determined by a neural network.
\Eq{
\pi^\theta_{T_i}(n) := \QxC{\tilde{N}^{\theta}_{T_i}=n}{s_{T_i},C^\theta_{T_{i-1}}} = \phi(n;N^\theta_{T_i},\xi_i)
}%
where the variances $\xi_i$ are added to the set $\theta$ of parameters which are subject to optimization. Policies are identified by the PPO algorithm with these densities. 

If we wish to limit the allowed strategies only to the bang-bang ones, we can simply restrict the action space to a discrete set. Hence, the neural network will directly return the vector of probabilities of each single admissible action. At the end of the training phase, the candidate optimal agent will take as $N^\theta_{T_i}$ the action with maximum probability as determined by the network.

Starting from the swing option control problem, we can define the action-value function if a specific action is taken at time $T_i$ as
\Eq{
\tilde{Q}^\theta_{T_i}(n) := \ExCo{ r_{T_i}(n) + \tilde{Q}^\theta_{T_{i+1}}(\tilde{N}^{\theta}_{T_{i+1}}) D(T_i,T_{i+1}) }{F_{T_i},C_{T_{i-1}}}
}%
where $r_{T_i}(n) := n ( F_{T_i} - K )$ is the reward at time $T_i$, and the filtration is extended to incorporate also the uncertainty in the actions.

If the action at time $T_i$ is integrated over all the possible choices, we can write the value function as
\Eq{
\tilde{V}^\theta_{T_i} := \ExCo{ r_{T_i}(\tilde{N}^{\theta}_{T_i})  + \tilde{V}^\theta_{T_{i+1}} D(T_i,T_{i+1}) }{F_{T_i},C_{T_{i-1}}}
}%

The PPO algorithm acts on $\theta$ to increase the value of an objective function which is made up of two main components, $L^A$ and $L^V$. The first component $L^A$ measures the goodness of the policy, and is related to the so-called advantage
\Eq{
A^{\theta}_{T_i} := \tilde{Q}^{\theta}_{T_i}(\tilde{N}^{\theta}_{T_i}) - \tilde{V}^{\theta}_{T_i}
}%
which has the property that the gradient of the expected reward equals 
\Eq{
	\nabla_{\theta} \left.\Exo{ A^{\bar{\theta}}_{T_i} \frac{\pi^{\theta}_{T_i}(\tilde{N}^{\bar{\theta}}_{T_i})}{\pi^{\bar{\theta}}_{T_i}(\tilde{N}^{\bar{\theta}}_{T_i})}}\right\rvert_{\bar{\theta}=\theta}
}
In practice one substitutes the unknown $A^\theta$ with a pathwise quantity which gives (approximately) the same gradient, namely
\Eq{
\hat{A}^\theta_{i} := \sum_{l=0}^{n_f-i-1} D(T_i,T_{i+l})\lambda^l \left[ r_{T_{i+l}}(\tilde{N}^{\theta}_{T_{i+l}}) + D(T_{i+l},T_{i+l+1}) V^{\theta}_{t+l+1} - V^{\theta}_{t+l} \right]
}%
Note that if $\lambda = 1$ then $\hat{A}^\theta_{T_i}$ telescopically reduces to the sum of realized discounted rewards, while if $\lambda < 1$ some bias is introduced by reducing the impact on $L^A$ of a rewards which are far in the future. This is done to get lower variance. For details, see \cite{Schulman2016}.

Instead, the second component $L^V$ of the objective function measures how well $V^{\theta}$ represents the value function $\tilde{V}^{\theta}$ of the policy $\pi^\theta$.

\begin{figure}
\begin{center}
\begin{tikzpicture}[node distance = 3em, auto, thick]
  \node [bluebox] (A1) {};
  \node [yellowbox, above left of=A1, xshift=-2em] (S1) {};
  \node [yellowbox, above left of=A1] (S2) {};
  \node [above left of=A1, xshift=2em, label={[xshift=0.17em, yshift=-1.03em]$\cdots$}] (S3) {};
  \node [yellowbox, above left of=A1, xshift=4em] (S4) {};
  \node [yellowbox, above left of=A1, xshift=6em] (S5) {};
  \node [redbox, below left of=A1, xshift=-2em] (R1) {};
  \node [redbox, below left of=A1] (R2) {};
  \node [below left of=A1, xshift=2em, label={[xshift=0.17em, yshift=-1.03em]$\cdots$}] (R3) {};
  \node [redbox, below left of=A1, xshift=4em] (R4) {};
  \node [redbox, below left of=A1, xshift=6em] (R5) {};
  
  \draw [fixto] (S1) -- (A1);
  \draw [fixto] (S2) -- (A1);
  \draw [fixto] (S4) -- (A1);
  \draw [fixto] (S5) -- (A1);
  \draw [fixto] (R1) -- (A1);
  \draw [fixto] (R2) -- (A1);
  \draw [fixto] (R4) -- (A1);
  \draw [fixto] (R5) -- (A1);

  \begin{scope}[on background layer]
    \node [draw=none, above of=S1] (T1u) {};
    \node [draw=none, below of=R1] (T1d) {};
    \node [draw=none, above of=S5] (T5u) {};
    \node [draw=none, below of=R5] (T5d) {};
    \draw (T1u) -- node [below right, near end, yshift=-1em] {$T_1$} (T1d);
    \draw (T5u) -- node [below right, near end, yshift=-1em] {$T_{n_f}$} (T5d);
  \end{scope}
  
  \node [bluebox, right=of A1, xshift=4em] (OU) {\shortstack{Policy and Advantage\\{\small\{$\pi^\theta_t,A^\theta_t$\}}}};
  \draw [floatto, draw=mydarkblue!75!black!100] (A1) -- (OU);
  
\end{tikzpicture}
\caption{An episode is a simulation of the state up to the swing option maturity. The agent interacting with the environment calculates on each fixing date $t$ the policy density and the advantage for selecting an action $\tilde{N}_{T_i}^{\theta}$ at such time.}
\label{fig:ppo}
\end{center}
\end{figure}
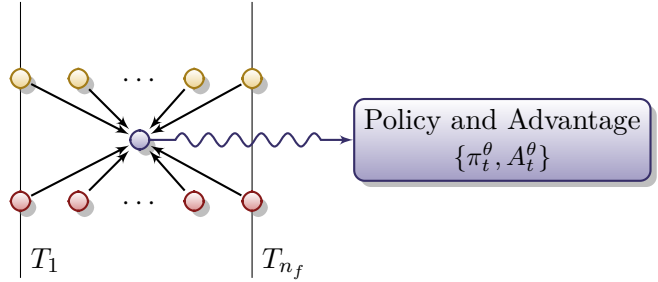

We illustrate in Figure~\ref{fig:ppo} the PPO algorithm. Each PPO batch is formed by episodes in which  the state is simulated up to the swing option maturity. The agent interacting with the environment calculates on each fixing date $t$ the policy density and the advantage for selecting an action $\tilde{N}_t^{\theta}$ at such time.

After the sampling process a new policy is proposed using a Stochastic Gradient Descent (SGD) with respect to the $\theta$ parameters:
\Eq{
\theta_{k+1} = \theta_k - \rho \cdot \Exo{ \nabla_{\theta} \left.\sum_{i=1}^{n_f} \left( L^A_{T_i}(\theta) - \beta L^V_{T_i}(\theta)\right)\right\rvert_{\theta=\theta_k}}
}%
\Eq{
L^A_{T_i}(\theta) := \min\left\{  \frac{\pi^{\theta}_{T_i}(\tilde{N}^{\theta_k}_{T_i})}{\pi^{\theta_k}_{T_i}(\tilde{N}^{\theta_k}_{T_i})} \hat{A}^{\theta_k}_{i}, \clip{1-\varepsilon}{\frac{\pi^{\theta}_{T_i}(\tilde{N}^{\theta_k}_{T_i})}{\pi^{\theta_k}_{T_i}(\tilde{N}^{\theta_k}_{T_i})}}{1+\varepsilon} \!\hat{A}^{\theta_k}_{i} \right\}
}%
\Eq{
L^V_{T_i}(\theta) := \left(  V^\theta_{T_i} - (\hat{A}^{\theta_k}_{i} +  V^{\theta_k}_{T_i}) \right)^{\!2}
}%
where the expected value is estimated over a batch of episodes, $\rho$ is a learning rate, $\clip{a}{x}{b}$ is the clip function (capped and floored linear function), while $\varepsilon$ and $\beta$ are hyper-parameters.

One of the key ideas of PPO is to ensure that a new policy update is ``close'' to the previous policy by clipping the advantages. The ratio behind this choice is to keep the new policy $\pi^{\theta_{k+1}}$ within a neighbourhood of the old one $\pi^{\theta_k}$ where one can trust both the first order approximation to the objective function given by the stochastic gradient, and the function $V^{\theta_k}$ used in the estimation of the advantage. Once the policy is updated, the experiences are thrown away and a newer batch is collected with the newly updated policy.

\subsection{Numerical Investigations with PPO}
\label{sec:numerics-ppo}

We focus in this example on a swing option contract with at-the-money fixed strike. The mean reversion is equal to $1$ in all experiments.

Several PPO hyper-parameters will be kept fixed to the following values: $\lambda=0.95$, $\varepsilon=0.2$, $\rho=0.0003$. These are general purpose defaults which are proposed in the original paper \cite{Schulman2017} and/or hard-coded in the baselines implementation. The trainable parameters $\theta$ are updated once every $2048$ training episodes.

In all experiments, two distinct feed-forward neural networks with hyperbolic tangent activation function are used to compute respectively the action and the value function at all times $t$, where the network inputs are $T_i$ expressed as a year fraction, the total consumption to-date remapped linearly at each time so that its domain is always $[-0.5, 0.5]$, and $\log (F_{T_i} / F_{T_0})$. In this way all inputs are well normalized, which helps the training of the network. The output layer is linear, i.e.\ no activation function is applied.

Both when the hypotheses which guarantee the existence of bang-bang optima are satisfied, and when they are not, we can allow for general $[N_m,N_M]$-valued actions; in the former case, the learning algorithm should find out by itself that the best strategy only involves bang-bang consumptions. The continuous-valued consumption is obtained by clipping the network's output to the interval $[0,1]$ and then remapping the result linearly so that $0$ and $1$ correspond respectively to the minimum and maximum admissible consumptions given both daily and global constraints. When we want to force bang-bang strategies instead, then only the minimum and maximum are considered as admissible actions, and a softmax layer remaps to probabilities the output units corresponding to these two actions.

\subsubsection{Neural Network Fine Tuning}

We explored several possible architectures of the neural network to investigate whether it impacts the final price and/or the number of iterations required for convergence. To this aim, we considered an option with a comparatively short delivery period of one week, and constraints 
\Eq{
N_m = 0 \,{\rm MWh}
\;,\quad
N_M = 1 \,{\rm MWh}
}
\Eq{
C_m = 3 \,{\rm MWh}
\;,\quad
C_M = 5 \,{\rm MWh}
}%
On this payoff, we tried both wide and deep architectures: 1 hidden layer of with 64 units (wide), 5 hidden layers with 4 units each (deep), and 5 hidden layers with 64 neurons each (wide and deep).

\begin{figure}
\begin{center}
\includegraphics[scale = 0.7]{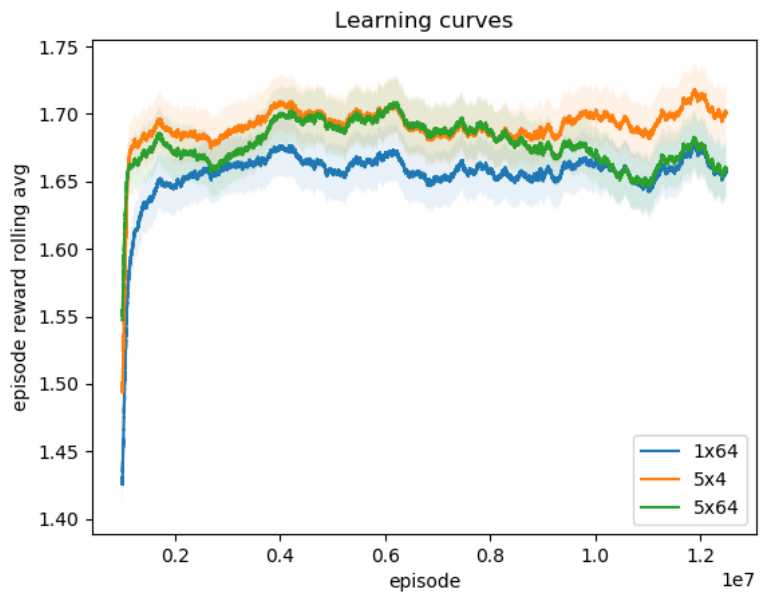}
\end{center}
\caption{Learning curves. On the horizontal axis the number of training episodes. The solid lines are the moving average of the realized rewards on the last $10^6$ episodes. The shadows represent the 98\% confidence intervals.}
\label{fig:learning}
\end{figure}

We show in Figure~\ref{fig:learning} the learning curves of the PPO algorithm with $\beta$ fixed to $0.5$ (the default in the baselines implementation). We can see that the shallow architecture finds a slightly suboptimal policy regardless of the high number of units (and hence of parameters). Deeper networks work better, but the very complex 5-by-64 network is prone to over-fitting, and indeed its performance deteriorates if optimized for too long. Hence, in what follows we focus on the 5-by-4 network.

\subsubsection{Comparison with the LSMC Algorithm}

In this section, we consider the contracts with maturity of one month delivering in May 2018 which were analysed in sections \ref{sec:lsmc} and \ref{sec: bang-bang} in the context of LSMC pricing.

\begin{table}
\begin{center}
\begin{tabular}{|r|c c|}
    \hline
                          & {All Strategies}    &  {Only Bang-Bang} \\ \hline
    {Out of Theorem Hyp.} & $7.92 \pm 0.04$ & $7.74 \pm 0.04$ \\ \hline
    {Within Theorem Hyp.} & $8.40 \pm 0.04$ & $8.37 \pm 0.04$ \\ \hline
\end{tabular}
\end{center}
\caption{Swing option prices in four different cases, obtained with RL. We consider the reference case with $C_m=12.5\,{\rm MWh}$ not satisfying the Theorem 2 in \cite{Bardou2009}, and a variant by changing the constraint to $C_m=12\,{\rm MWh}$ so that the Theorem is satisfied. Prices are calculated either allowing all possible strategies or only the bang-bang ones. One-sigma statistical errors are displayed.}
\label{tab:compare_1m}
\end{table}

After training, the option can be priced using either the LSMC or the RL candidate optimal policy. We therefore run a Monte Carlo simulation with 1 million paths to get the price.


We performed a grid search on the hyperparameter $\beta$ and found that $\beta=0.01$ was more effective than the default $\beta=0.5$, corresponding to slower updates of the network which approximates the value function. Moreover, since the objective function is not convex, we run each optimization four times with different random starting guesses for $\theta$, and then choose the optimized network with the best in-sample performance on the last 1,000,000 training episodes. The out-of-sample results of such network are shown in Table~\ref{tab:compare_1m}, and they are compatible with the LSMC results in Table~\ref{tab:bb} within statistical uncertainty.

\begin{figure}
\begin{center}
\includegraphics[trim={0 1cm 0 2.1cm},clip,scale=0.7]{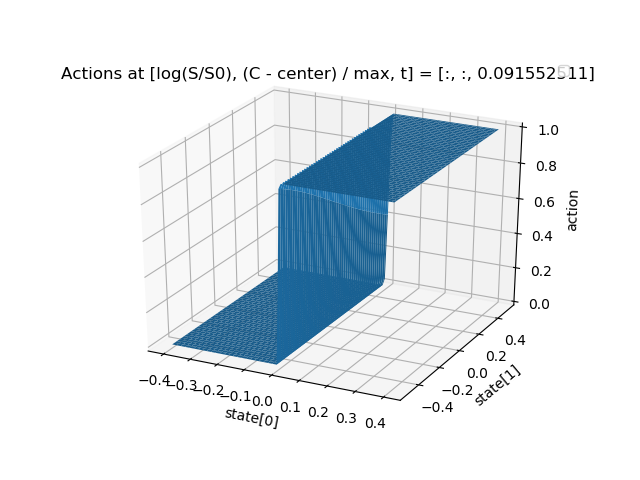}
\end{center}
\caption{Normalized consumption as a function of normalized log-spot and consumption, on the fourth decision date. The axes are: $\log (F_{T_i} / F_{T_0})$; total consumption to-date remapped linearly so that its domain is $[-0.5, 0.5]$; today's consumption remapped linearly so that its domain is $[0,1]$.}
\label{fig: action}
\end{figure}

We also see that the unconstrained PPO agent successfully identifies a strategy of bang-bang type for the case $C_m = 12\, {\rm MWh}$ in which we know that it is optimal to do so. This is exemplified by Figure~\ref{fig: action}, where we fix a decision time and plot the chosen action as a function of the other two coordinates of the network input (i.e.\ normalized log-spot and consumption).

\section{Conclusion and Further Developments}
\label{sec:conclsion}

In this paper we presented a new model to price swing option contracts. The model is able to calibrate liquid market quotes and to imply the volatility smile for futures contracts with different delivery periods. We show also how to extend the model to include spikes into its dynamics. The pricing algorithm is implemented both by using a least-square Monte Carlo approach and by means of recent reinforcement learning algorithms, such as the proximal policy optimization algorithm. Using the former, we investigate option prices and optimal strategies for different configuration of the model, and we test the impact of constraining the choice of the control problem only to bang-bang strategies. The aim of exploring techniques based on reinforcement learning is due to the fact that we wish to investigate calculation tools more suitable in high-dimensional settings. We find that this novel techniques also gives accurate results. This paper focuses on situations where other techniques are available as a benchmark, to gather evidence on the robustness of the approach; we leave for future developments the exploration of settings where it could be the only possibility.

\newpage

\bibliographystyle{plainnat}
\bibliography{cosmile}

\end{document}